\documentclass{article}
\usepackage{amssymb}

\usepackage[preprint]{neurips_2026}

\usepackage[utf8]{inputenc} 
\usepackage[T1]{fontenc}    
\usepackage{hyperref}       
\usepackage{url}            
\usepackage{booktabs}       
\usepackage{amsfonts}       
\usepackage{nicefrac}       
\usepackage{microtype}      
\usepackage{xcolor}         

\usepackage{multirow}
\usepackage{amssymb}
\usepackage{array}
\usepackage{graphicx} 
\usepackage{wrapfig}
\usepackage{pifont}

\usepackage{amsmath}
\usepackage{mathtools}
\usepackage{amsthm}
\usepackage{siunitx}
\usepackage[most]{tcolorbox}

\newcommand\shortsection[1]{\vspace{.75ex}{\noindent\bf #1.}}

\title{How to Steal Reasoning Without Reasoning Traces}

\author{%
  Tingwei Zhang \quad John X. Morris \quad Vitaly Shmatikov \\
  Department of Computer Science\\
  Cornell Tech
}

\begin{document}

\maketitle

\begin{abstract}
  Many large language models (LLMs) use reasoning to generate responses but do not reveal their full reasoning traces (a.k.a.\ chains of thought), instead outputting only final answers and brief reasoning summaries.  To demonstrate that hiding reasoning traces does not prevent users from ``stealing'' a model's reasoning capabilities, we introduce \emph{Trace Inversion} models that, given only the inputs, answers, and (optionally) reasoning summaries exposed by a target model, generate detailed, synthetic reasoning traces.  We show that (1) traces synthesized by Trace Inversion have high overlap with the ground-truth reasoning traces (when available), and (2) fine-tuning student models on inverted traces substantially improves their reasoning and enables distillation from proprietary, black-box LLMs.  
\end{abstract}

\section{Introduction}
\label{sec:intro}
Large language models (LLMs) with explicit reasoning capabilities perform well on math, coding, and scientific analysis tasks.  Trained via supervised fine-tuning, instruction tuning, and reinforcement learning, these models generate multi-step internal reasoning traces, aka chains of thought, leading to final answers \citep{wei2022chain, ouyang2022training, chen2025skip}. 
At inference time, these traces decompose complex problems into intermediate steps.  During training, they help with capability transfer.  Fine-tuning a student model on step-by-step reasoning traces generated by a teacher model can transfer much of the latter's reasoning capability, substantially outperforming fine-tuning on final answers alone \citep{hinton2015distilling, gou2021knowledge, shridhar2023distilling, chen2025skip}. 

Exposing full chains of thought is a potential intellectual property risk.  For example, in February 2026 Anthropic accused several competing labs of large-scale distillation campaigns that allegedly elicited data from Anthropic's reasoning models and used it for training~\citep{anthropic_distillation_2026}.  Furthermore, reasoning traces may encode sensitive system prompts, safety policies, or private contextual information \citep{green2025leaky, deepseek_exploit, zhou2025hidden}.
To mitigate these risks,  commercial APIs typically reveal only a short reasoning summary and the final answer \citep{openai_reasoning_docs, gemini_thoughts_docs, anthropic_extended_thinking}.  This assumes that exposing summaries in lieu of detailed chains of thought preserves the practical benefits of reasoning models while preventing capability stealing.

\begin{figure*}[t]
  \centering  
  \includegraphics[width=0.93\linewidth]{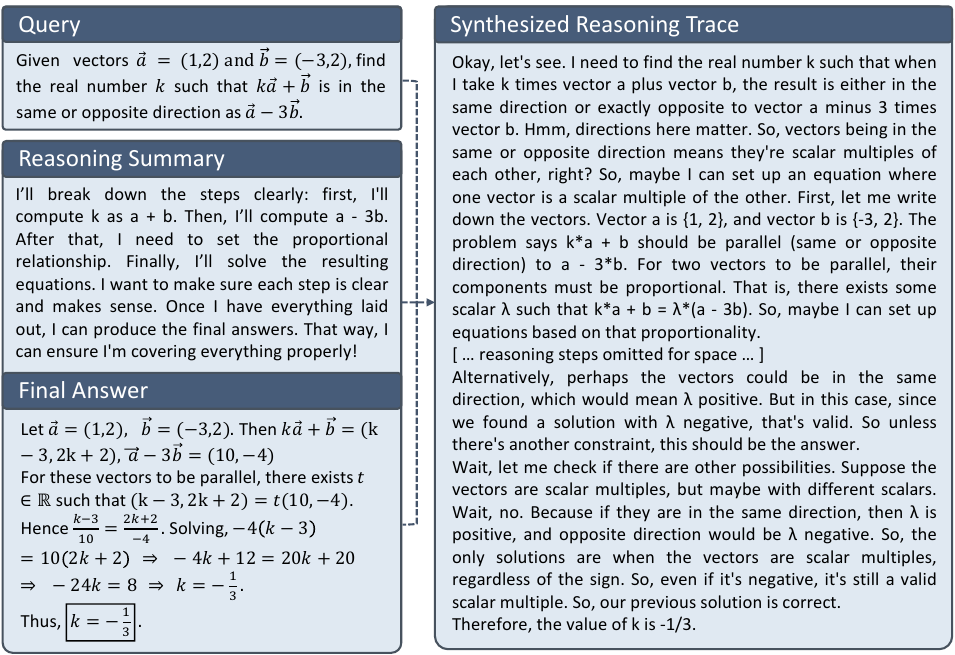} 
  \caption{\textbf{Example of Trace Inversion against \textbf{GPT-5.4 mini}.} Given only the input, final answer, and reasoning summary from a commercial black-box model, Trace Inversion synthesizes a detailed reasoning trace that can be used for supervised fine-tuning of a student model.}
  \label{fig:example_1}
  \vskip -0.1in
\end{figure*}

In this paper, we challenge this assumption. We introduce \textbf{Trace Inversion}, a framework for synthesizing detailed reasoning traces from only the final answers and, optionally, compressed summaries (Figure~\ref{fig:example_1}) exposed by reasoning models. Our inversion models synthesize traces that are similar to the ground truth (e.g., 81\% token length recovery and 52.76 token-overlap F1 score for DeepSeek-R1 traces, even though inversion is trained on a much weaker surrogate model).  Most importantly, \textbf{inverted traces can ``teach'' a student model how to reason even when the black-box teacher model does not reveal its reasoning and is stronger than either the student, or the inversion model}.  Across math, scientific reasoning, and coding benchmarks, models fine-tuned on inverted traces outperform those fine-tuned on (a) surrogate reasoning traces, or (b) other models' answers and reasoning summaries.  This demonstrates that hiding chains of thought does not prevent users from stealing the model's reasoning capability.

For example, fine-tuning \texttt{Qwen-2.5-7B-Instruct} on traces inverted from the answers and summaries of \texttt{gpt-5.4-mini-2026-03-17} (GPT-5.4 mini), a much stronger, commercial, black-box model, 
by (weak) \texttt{DeepSeek-R1-Distill-Qwen-1.5B}, reaches 31.6\% on JEEBench, vs.\ 19.7\% when fine-tuned on \texttt{R1-Distill}'s own reasoning traces. 
Fine-tuning \texttt{Llama-3.1-8B-Instruct} on traces inverted from GPT-5.4 mini by an inversion model based on \texttt{Deepseek-R1} reaches 52.4\% on MATH500, vs.\ 16.4\% when fine-tuned on the GPT-5.4 mini's reasoning summaries and answers.  In some cases, fine-tuning on inverted traces yields even better performance than the ground-truth traces (the latter can contain backtracking and dead ends, whereas inverted traces are typically clean forward reasoning).

To facilitate research on model-stealing attacks and defenses, we release our code. \footnote{\url{https://github.com/Tingwei-Zhang/Trace_Inversion_Attack}}


\section{Background and Related Work}

\textbf{Reasoning models.}
Many modern large language models rely on explicit multi-step reasoning.  Chain-of-thought (CoT) prompting showed that eliciting intermediate reasoning steps substantially improves performance on arithmetic, logical, and symbolic benchmarks \citep{wei2022chain,kojima2022large}.  Subsequent work demonstrated improved accuracy and robustness when models are guided to produce coherent intermediate steps and not just final answers \citep{wang2022self}.
Instruction tuning and reinforcement learning (RL) can incorporate reasoning-oriented objectives \citep{ouyang2022training,bai2022training}, e.g., in models like DeepSeek-R1 \citep{guo2025deepseek} and OpenAI’s o-series \citep{openai_o1_system_card}.  Related models such as QwQ-32B and s1 further show that RL can enable smaller models to approach the reasoning performance of large base models \citep{qwen2024qwq,muennighoff2025s1}.

\textbf{Information leakage from language models.}
Black-box language models can leak information about training data and inference-time inputs through their outputs.  This includes training-data extraction, where a model reveals memorized training inputs through its generations \citep{carlini2021extracting}, and membership inference \citep{shokri2017membership}, which determines whether specific data points were included in the model's training data, including in-context learning settings \citep{mattern2023membership,wen2024membership}. 

A separate line of work studies leakage of inference-time inputs.  Model inversion infers information about the inputs given just the outputs of black-box models \citep{morris2024language}.  Prompt inversion is a special case where only part of the input (e.g., system prompt) is secret and can be extracted via adversarial queries \citep{zhang2024effective}, or prompt inversion methods \citep{zhang2024extracting,yang2024prsa}. 

\textbf{Model distillation and capability stealing.}
Model distillation transfers capabilities from a teacher model to a student model. In cooperative settings, the student has full or partial access to the teacher and aims to match its outputs or predictive distributions \citep{hinton2015distilling}. In adversarial settings, an attacker seeks to transfer capabilities from a deployed model using limited external access.  This \emph{capability stealing} was originally studied for image classifiers \citep{tramer2016stealing,orekondy2019knockoff} and extended to language models \citep{krishna2019thieves, wallace2020imitation}.  The attack is typically performed under black-box access, where the attacker observes only the model's outputs and has no access to its internal states, parameters, or training data.

The effectiveness of distillation and capability stealing depends strongly on the information available from the teacher model.  Training student models on teacher-generated reasoning traces (rather than final answers alone) leads to substantial gains on reasoning tasks \citep{wei2022chain,shridhar2023distilling}. Variants include partial reasoning supervision or skipping intermediate steps to balance cost and performance \citep{chen2025skip}.  We focus on the realistic setting where the teacher's reasoning traces are \textbf{not} available from the model API.

\section{Threat Model}
\label{sec:threat-model}

\shortsection{Problem setting}
Consider a black-box \emph{victim model} \(V\) that, given an input \(x \in \mathcal{X}\), internally generates a reasoning trace \(t \in \mathcal{T}\) that leads to the
final answer \(y \in \mathcal{Y}\).  The user receives only $y$ and, optionally, a short reasoning summary or ``bubble'' \(b^{\star} = C(t) \in \mathcal{B}\) such that \(|b^{\star}| \ll |t|\).  This setting corresponds to the APIs of commercial reasoning models.

\shortsection{Attacker’s goal}
The attacker’s primary objective is to improve the reasoning capabilities of their own model \(S\) by exploiting black-box access to a stronger reasoning model \(V\), without direct access to \(V\)’s internal reasoning and using only the publicly available resources listed below. To this end, given tuples \((x, y)\) or \((x, b^{\star}, y)\) output by $V$, the attacker aims to synthesize reasoning traces \(\hat{t}\) that are logically consistent with the observed tuples.  Whether they match the victim’s true internal reasoning or not, their purpose is to support effective fine-tuning of \(S\) on \((x, \hat{t}, y)\).

\shortsection{Attacker's auxiliary models}
The attacker may use its own surrogate reasoning model \(V'\), compression model \(C'\), inversion model \(I\), and student model \(S\).  These models may be public or private to the attacker.  The only requirement is that the attacker be able to fine-tune the inversion model \(I\) and student model \(S\).   The \textbf{surrogate reasoning model} \(V'\) shoudl produce complete reasoning traces \((t', y')\), which can be used to train the inversion model.  Typically, \(V'\) is much weaker than the victim model (e.g., \texttt{R1-distill} vs.\ GPT-5.4 mini in our experiments).

We do \emph{not} assume that the attacker has access to the true compression method $C$ used by the victim $V$ (e.g., it is not publicly known how GPT models produce reasoning summaries).  To ``emulate'' summarization, the attacker employs a \textbf{reasoning compression model} \(C'\) that maps surrogate traces \(t'\) to summaries \(b' = C'(t'; \pi)\) under a fixed prompt template \(\pi\) without additional fine-tuning.  The attacker then trains an \textbf{inversion model} \(I\) to synthesize reasoning traces from $V$’s observable input–output pairs \((x, y)\) and, when available, summaries \(b^{\star}\), and uses these traces to fine-tune a \(S\).

\noindent
\textbf{Public reasoning datasets.}  
The attacker can draw on public datasets \(D\) of diverse reasoning-task inputs \(x\) (e.g., math, code, and logic questions), such as \texttt{OpenThoughts-114k} \citep{guha2025openthoughtsdatarecipesreasoning}.  
We use them as a source of inputs for (1) querying the surrogate reasoning model to obtain its traces, and (2) querying the victim model to obtain answers and summaries that can be fed into the Trace Inversion model.

\section{Inversion Methodology}
\label{sec:method}

\begin{figure*}[t]
  \centering  
  \includegraphics[width=1.0\linewidth]{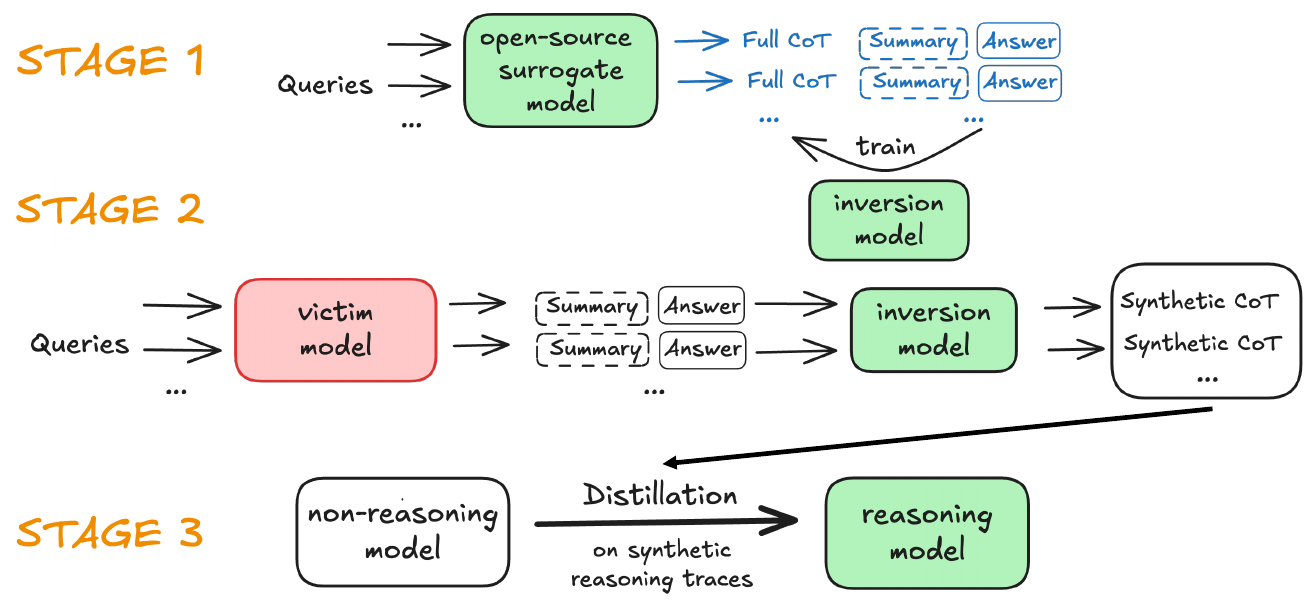}
  \caption{\textbf{Trace Inversion pipeline.}}
  \label{fig:inversion_model}
\end{figure*}

\emph{Trace Inversion} is a three-stage attack pipeline (Figure~\ref{fig:inversion_model}).
We consider a \emph{summary setting}, where the victim model exposes a short reasoning summary, and a stricter \emph{no-summary setting}, where only the final answer is available.  Each setting requires a separate inversion model.

\subsection{Stage 1: Training the Inversion Model}
\label{sec:method-train-inverter}

\noindent\textbf{Training data.}
The attacker samples inputs \(x'\) from public reasoning datasets, submits them to the surrogate model 
\(V'\), and obtains traces $t'$ and answers $y'$, yielding dataset
$\mathcal{D}_1 = \{(x', t', y')\}$.  To emulate commercial APIs, each surrogate trace \(t'\) is optionally compressed into a summary \(b'=C'(t'; \pi)\) using model \(C'\) (see Appendix \ref{box:summarization_prompt}). This yields datasets
$\mathcal{D}_2^{\text{sum}} = \{(x', b', y', t')\}$ and $\mathcal{D}_2^{\text{nosum}} = \{(x', y', t')\}$ for the summary and no-summary settings, respectively. \textbf{We use our own compression model because the actual summarization method used by the black-box victim model is not known} and thus cannot be used to (trace, summary) pairs we need for training.

\noindent\textbf{Training objective.}
In the summary setting, the inversion model \(I_{\text{sum}}(x, y, b) = \hat{t}\) is trained to maximize the likelihood that its output matches the true trace $t'$:
\begin{equation}
  \mathcal{L}_{\mathrm{inv}}^{\text{sum}}
=
\mathbb{E}_{(x', y', b', t') \sim \mathcal{D}_2^{\text{sum}}}
\left[
-\log p_{I_{\text{sum}}}\!\big(t' \mid x', y', b'\big)
\right]
\end{equation}
The no-summary setting is similar: $I_{\text{nosum}}(x, y) = \hat{t}$ and the training objective is:
\begin{equation}
  \mathcal{L}_{\mathrm{inv}}^{\text{nosum}}
=
\mathbb{E}_{(x', y', t') \sim \mathcal{D}_2^{\text{nosum}}}
\left[
-\log p_{I_{\text{nosum}}}\!\big(t' \mid x', y'\big)
\right]
\end{equation}

Both objectives are implemented using teacher forcing with token-level cross-entropy.

\subsection{Stage 2: Inverting Victim's Outputs}
\label{sec:method-invert-victim}
The attacker first queries the victim $V$ on inputs $x$, collecting tuples $(x, y, b^{\star})$ (or $(x, y)$ in the no-summary setting). The trained inversion model is then applied to the tuples: $\hat{t} = I_{\text{sum}}(x, y, b^{\star})$ or $\hat{t} = I_{\text{nosum}}(x, y)$, producing synthetic reasoning traces $\hat{t}$ that are compatible with $V$'s observed outputs and can serve as approximations of its hidden internal reasoning in the next stage.

\subsection{Stage 3: Student Distillation}
\label{sec:method-distillation}

In the final stage, the attacker distills the approximation of the victim’s reasoning synthesized by the inversion model into the student model \(S\). Each training example consists of an input \(x\) and the synthetic reasoning trace \(\hat{t}\) concatenated with the answer \(y\) into a single supervision target
$y^{+} = [\,\hat{t};\, y\,]$.  \(S\)
is fine-tuned using standard teacher-forced cross-entropy:
\begin{equation}
  \mathcal{L}_{\mathrm{student}}
=
\mathbb{E}_{(x, \hat{t}, y)}
\left[
-\log p_S\!\big(y^{+} \mid x\big)
\right]
\end{equation}

\noindent\textbf{Optional: augmenting with surrogate data.} Since the attacker already generated the surrogate dataset $\mathcal{D}_1 = \{(x', t', y')\}$ for training the inversion model, these examples can be added to the distillation data at no additional cost. The student can be fine-tuned on $\{(x, \hat{t}, y)\} \cup \mathcal{D}_1$ either as a single mixed corpus, or in a curriculum (surrogate traces first, inverted traces second). We report this as an ablation in \S\ref{sec:eval-gpt-stealing}; the default pipeline uses only the inverted traces.

\section{Evaluation}
\label{sec:eval}

\subsection{Experimental Setup}
\label{sec:eval-setup}

\paragraph{Datasets.}
\texttt{OpenThoughts-114k} (OpenThoughts) is a public reasoning dataset spanning math, science, programming, and logic. Each example is an input $x$, an R1-generated reasoning trace, and the final answer.  We sample two disjoint splits of 10k inputs each: a \emph{surrogate split} for querying an open-source reasoning model to obtain $(x',t',y')$ used to train inversion models, and a \emph{victim split} for querying the victim to obtain $(x,t,y)$ (open-weight victim) or $(x,b^\star,y)$ (black-box victim).

\noindent
\textbf{Models.}
As surrogate models, we
use R1 or \texttt{DeepSeek-R1-Distill-Qwen-1.5B}\footnote{\url{https://huggingface.co/deepseek-ai/DeepSeek-R1-Distill-Qwen-1.5B}}, which we refer to as \textbf{R1-Weak} to emphasize that it is weaker than the victim models. We use R1 as the open-weight victim, and \texttt{gpt-5.4-mini-2026-03-17} (GPT-5.4 mini) as the black-box victim, both accessed via their commercial APIs.
We use \texttt{Qwen2.5-7B-Instruct} (Qwen) as (1) a trace summarization model, and (2) backbone for our for inversion models.  To evaluate utility of inverted traces, we fine-tune two student models: \texttt{Qwen2.5-7B-Instruct} (Qwen) and Llama-3.1-8B-Instruct (Llama).

\noindent
\textbf{Inversion metrics.}
When the victim is open-weight (i.e., R1), we compare the synthesized trace $\hat{t}$ to the ground-truth trace $t$ using \emph{Len} (token count; a coarse proxy for completeness), \emph{BLEU} ($n$-gram precision, capturing surface lexical similarity), \emph{TF1} (token-overlap F1, i.e., the harmonic mean of token-level precision and recall), and \emph{ROUGE-1/2/L} (unigram, bigram, and longest-common-subsequence overlap). As mentioned above, these metrics are auxiliary diagnostics. Our main evaluation metric is the reasoning performance of student models fine-tuned on inverted traces.

\noindent
\textbf{Fine-tuning data for student models.}
We consider five options: \emph{Answer-only} (victim's answers), \emph{Summary+Answer} (victim's summaries and answers), \emph{Surrogate-Trace} (surrogate's reasoning traces and answers), \emph{Synthesized-Trace (ours)} (our inverted traces and victim's answers), and \emph{Victim-Trace (oracle)} (victim's actual traces and answers\textemdash not available in practice).

\noindent
\textbf{Reasoning benchmarks.}
To evaluate whether fine-tuning student models on inverted traces improves their reasoning performance, we use three benchmarks: \emph{MATH500} \cite{hendrycks2021measuring} evaluates multi-step mathematical problem solving on competition-level questions; \emph{JEEBench} \cite{arora2023have} measures structured problem solving across physics, chemistry, and mathematics; \emph{LiveCodeBench} (LCB) \cite{jain2025livecodebench} evaluates code generation and debugging using execution-based metrics.

\begin{wraptable}{r}{0.54\textwidth}
  \vspace{-1.5\baselineskip}
  \centering
  \setlength{\belowcaptionskip}{8pt}
  \caption{\textbf{Feature distribution of our surrogate summaries vs.\ GPT-5.4 mini.}.}
  \label{tab:stylematch}
  \small
  \setlength{\tabcolsep}{1.2pt}
  \begin{tabular}{lcc}
    \toprule
    & Ours ($C'$ on R1 traces) & GPT-5.4 mini \\
    \midrule
    Median tokens & 537 & 592 \\
    Bold-header sections & 94.1\% & 92.9\% \\
    First-person prose & 97.3\% & 97.0\% \\
    LaTeX & 79.1\% & 71.9\% \\
    \bottomrule
  \end{tabular}
\end{wraptable}

\noindent
\textbf{Hyperparameters.}
All models are fine-tuned using 
\texttt{num\_train\_epochs = 3},
\texttt{learning\_rate = 1e-5},
\texttt{warmup\_ratio = 0.1},
\texttt{cutoff\_len = 16384}.
Unless otherwise specified, we use each model's the default optimizer and tokenizer and follow the original evaluation protocols of each benchmark.  We runn training and evaluation on 8 NVIDIA Tesla A100s (80GB).

\subsection{Fidelity of Trace Inversion}
\label{sec:eval-inversion}

We first evaluate whether the inversion model synthesizes traces that resemble the victim's true reasoning.  Against an open-weight victim (R1), we measure this directly.
Because GPT-5.4 mini does not reveal its traces, we match the distribution of summaries instead.

\shortsection{Matching summaries (GPT-5.4 mini)}
The inversion model is trained on the surrogate summaries $b' = C'(t'; \pi)$ and applied to the victim's summaries $b^{\star}$; it generalizes best when the two distributions match.  While GPT-5.4 mini hides its internal traces, its API reports the total number of reasoning tokens (for billing). We recover the trace length by subtracting visible-output tokens. Non-empty GPT-5.4 mini summaries have median length of $\approx$592 tokens and median compression of $\approx$$4.2\times$. Stylistically, they are first-person prose with bold-header sections and frequent LaTeX (based on lightweight regex heuristics).  We designed our compression prompt $\pi$ (Appendix~\ref{box:summarization_prompt}) to produce summaries that match this style.  Applied to the surrogate traces (from R1), the resulting summaries align with GPT-5.4 mini's summaries on median length and all dominant style features (Table~\ref{tab:stylematch}).

\shortsection{Matching reasoning traces (R1)}
Table~\ref{tab:inversion} reports how well traces inverted from R1's outputs match the ground truth.  When the inversion model is trained on traces from R1 itself, the training and evaluation distributions are the same, giving an upper bound on achievable fidelity: TF1 58.00 / 57.08 and R-L 29.01 / 27.97 in the summary and no-summary settings, respectively.

\begin{table*}[t]
  \centering
  \setlength{\belowcaptionskip}{8pt}
  \caption{\textbf{Trace inversion fidelity (victim = R1).}
  We report synthesized trace length (Len) and overlap metrics between synthesized traces $\hat{t}$ and R1 ground-truth traces $t$.
  \textbf{Bold} rows use a weaker surrogate (R1-Weak), corresponding to a realistic attack. Higher is better, except trace length.}
  \label{tab:inversion}
  \small
  \begin{tabular}{lllrrrrrr}
    \toprule
    \textbf{Setting} & \textbf{Inversion} & \textbf{Surrogate} & \textbf{Len} & \textbf{BLEU} & \textbf{TF1} & \textbf{R-1} & \textbf{R-2} & \textbf{R-L} \\
    \midrule
    Summary & Qwen (0-shot) & -- & 978.5 & 3.04 & 35.36 & 26.77 & 15.73 & 17.60 \\
    Summary & Qwen (FT) & R1 & 5{,}767.3 & 26.64 & 64.42 & 66.84 & 40.51 & 29.17 \\
    Summary & Qwen (FT) & \textbf{R1-Weak} & 4{,}971.9 & 16.64 & 52.76 & 56.59 & 29.72 & 24.84 \\
    \midrule
    No-summary & Qwen (0-shot) & -- & 1{,}040.0 & 3.31 & 37.58 & 28.30 & 16.39 & 17.97 \\
    No-summary  & Qwen (FT) & R1 & 6{,}020.9 & 25.07 & 57.08 & 65.43 & 38.91 & 27.97 \\
    No-summary  & Qwen (FT) & \textbf{R1-Weak} & 5{,}434.0 & 11.52 & 49.01 & 47.45 & 21.85 & 21.73 \\
    \bottomrule
  \end{tabular}
  \vskip -0.1in
\end{table*}

\shortsection{Prompting alone is insufficient} As a zero-shot baseline, we prompt Qwen (Appendix~\ref{box:zero_shot_inversion_summary}) with $(x, b^\star, y)$ or $(x, y)$ to synthesize a trace.  In both settings, this produces short, weakly aligned traces: length collapses to $\approx$1{,}000 tokens (vs.\ R1's ground-truth average of 6{,}130.6), with weak overlap (TF1 35.36 / 37.58; R-L 17.60 / 17.97). See Figure~\ref{fig:example_2} in the appendix for an illustrative example.

\shortsection{Weaker surrogate is nearly as effective} Training inversion on the R1-Weak surrogate recovers most of the gap: synthesized traces are long (4{,}972 / 5{,}434 tokens, $\approx$81--89\% of R1's average) and overlap substantially with the ground truth (TF1 52.76 / 49.01, R-L 24.84 / 21.73).  The drop from R1 to R1-Weak is small relative to the jump from zero-shot to any inversion model.  This corresponds to a realistic attack where the surrogate is weaker than the victim (see Figure~\ref{fig:example_3} in the appendix).

\shortsection{Summary vs.\ no-summary} Overlap metrics are slightly higher in the summary setting. This understates the qualitative gap: summaries provide explicit cues that inversion can latch onto, improving coherence and step-level correctness in ways that overlap metrics may not fully capture.

\subsection{Stealing Reasoning Capabilities of an Open-Weight Model (R1)}
\label{sec:eval-r1-stealing}

We evaluate whether fine-tuning a student on inverted traces improves its downstream reasoning. Using R1 as the victim allows us compare against both attacker-accessible baselines (Answer-only, Summary+Answer) and the Victim-Trace oracle. Table~\ref{tab:r1_stealing} reports accuracy for both students in the summary ($x,b^\star,y$ available) and no-summary ($(x,y)$) settings.  R1-Weak serves as a realistic weak surrogate, while R1 trained on its own data provides an upper bound on inversion quality.

It is important to note that even though it does not have explicit reasoning, our base instruction-tuned Qwen model appears to have been optimized for the MATH500 benchmark.  This means that fine-tuning can \emph{degrade} its performance (71.2\%$\rightarrow$61.0\% when fine-tuning on answers).  This matters for our threat scenario.  The most obvious black-box attack\textemdash fine-tuning the student on the victim's answers and summaries\textemdash can damage the student rather than teach it how to reason.  This highlights the utility of inversion models: they turn the victim's outputs into a useful supervision signal.

\begin{table*}[t]
  \centering
  \setlength{\belowcaptionskip}{6pt}
  \caption{\textbf{Downstream accuracy (\%) after fine-tuning when attacking R1.} \textit{Victim-Trace (oracle)} is an upper bound not available to realistic attackers. Entries marked with $^{\star}$ outperform the oracle (explained in \S\ref{sec:eval-r1-stealing}).}
  \label{tab:r1_stealing}
  \small
  \setlength{\tabcolsep}{4.4pt}  \begin{tabular}{lll
  S[table-format=2.1] S[table-format=2.1]
  S[table-format=2.1] S[table-format=2.1]
  S[table-format=2.1] S[table-format=2.1]}
    \toprule
    \multirow{2.5}{*}{\textbf{Setting}} & \multirow{2.5}{*}{\textbf{Training Sources}} & \multirow{2.5}{*}{\textbf{Surrogate}}
    & \multicolumn{2}{c}{\textbf{MATH500}}
    & \multicolumn{2}{c}{\textbf{JEEBench}}
    & \multicolumn{2}{c}{\textbf{LiveCodeBench}} \\
    \cmidrule(lr){4-5} \cmidrule(lr){6-7} \cmidrule(lr){8-9}
    & & & \multicolumn{1}{c}{\textbf{Qwen}} & \multicolumn{1}{c}{\textbf{Llama}}
      & \multicolumn{1}{c}{\textbf{Qwen}} & \multicolumn{1}{c}{\textbf{Llama}}
      & \multicolumn{1}{c}{\textbf{Qwen}} & \multicolumn{1}{c}{\textbf{Llama}} \\
    \midrule
    -- & No finetuning & -- & 71.2 & 42.1 & 28.3 & 16.6 & 28.9 & 13.1 \\
    -- & Answer-only & -- & 61.0 & 44.0 & 21.6 & 17.2 & 25.4 & 9.6 \\
    -- & Surrogate-Trace & R1-Weak & 63.2 & 48.8 & 19.7 & 11.3 & 22.9 & 9.1 \\

    \midrule
    Summary & Summary+Answer & -- & 63.0 & 45.8 & 24.0 & 18.2 & 25.6 & 11.7 \\
    Summary & \textbf{Synthesized-Trace (ours)} & \textbf{R1-Weak} & 71.8 & 50.2 & 36.3 & 24.2\textsuperscript{*} & 30.9 & 8.8 \\
    Summary & Synthesized-Trace (ours) & R1 & 74.4\textsuperscript{*} & 59.2 & 37.1 & 26.1\textsuperscript{*} & 32.2 & 9.9\textsuperscript{*} \\
    \midrule
    No-sum. & \textbf{Synthesized-Trace (ours)} & \textbf{R1-Weak} & 66.4 & 44.8 & 29.7 & 16.3 & 26.2 & 9.2 \\
    No-sum. & Synthesized-Trace (ours) & R1 & 74.1\textsuperscript{*} & 62.0\textsuperscript{*} & 33.2 & 25.2\textsuperscript{*} & 29.1 & 8.6 \\
    \midrule
    -- & \textit{Victim-Trace (oracle)} & \textit{--} & \textit{72.2} & \textit{59.6} & \textit{43.7} & \textit{23.8} & \textit{33.2} & \textit{9.8} \\
    \bottomrule
  \end{tabular}
  \vskip -0.06in
\end{table*}

\noindent\textbf{Inverted traces beat every attacker-accessible baseline.} With the R1-Weak surrogate, Synthesized-Trace beats Answer-only, Summary+Answer, and Surrogate-Trace in almost every experiment. On MATH500, it lifts Qwen from 63.0\% (Summary+Answer) to 71.8\% and Llama from 45.8\% to 50.2\%; JEEBench gains are larger (Qwen 24.0\%$\rightarrow$36.3\%, Llama 18.2\%$\rightarrow$24.2\%). Improvement over Surrogate-Trace is critical: it shows that \textbf{fine-tuning on inverted traces outperforms fine-tuning on traces from a surrogate reasoning model.}

\noindent\textbf{No-summary inversion is still effective.} Removing $b^\star$ degrades performance but Synthesized-Trace still improves over Answer-only and Surrogate-Trace (e.g., Qwen JEEBench 19.7\%$\rightarrow$29.7\%). With a strong R1 surrogate, no-summary inversion produces traces that help Qwen reach 74.1\% on MATH500 and 33.2\% on JEEBench, close to the oracle on both.

\noindent\textbf{Inverted traces approach the oracle.} The Victim-Trace oracle is our upper bound. Inverted traces close most of the gap: with an R1-Weak surrogate in the summary setting, Qwen matches oracle JEEBench (36.3\% vs.\ 43.7\%) and trails MATH500 by 0.4 points (71.8\% vs.\ 72.2\%). Llama closes the JEEBench gap from 11.3\% (Surrogate-Trace) to 24.2\% vs.\ 23.1\% oracle.

\noindent\textbf{Inversion can outperform ground truth when surrogate and victim are the same model.} When the surrogate and the victim are the same model (R1, in our case), Synthesized-Trace can even beat Victim-Trace oracle: in the no-summary setting this happens on Qwen/MATH500 (74.1\% vs.\ 72.2\%), Llama/MATH500 (62.0\% vs.\ 59.6\%), and Llama/JEEBench (25.2\% vs.\ 23.8\%). R1's raw traces often include backtracking and dead ends, introducing noise into student fine-tuning.  Our inversion model,  conditioned on the final answer, produces clean forward reasoning, which ``denoises'' R1's reasoning and provides a better fine-tuning signal.
See Appendix Figure~\ref{fig:r1_vs_inverter_backtrack} for a side-by-side example.

\noindent\textbf{Weak surrogates hurt student training but help inversion training.} Directly distilling R1-Weak can degrade the student (e.g., Qwen MATH500 71.2\%$\rightarrow$63.2\%; LCB 28.9\%/13.1\%$\rightarrow$22.9\%/9.1\% for Qwen/Llama).  When the same traces are used only to train the inversion model, which is then applied to the stronger victim's outputs, inverted traces are much more effective for student training.  This explains why fine-tuning on inverted traces outperforms fine-tuning on surrogate traces.

\noindent\textbf{LiveCodeBench gains are weaker.} Llama's LCB accuracy drops on Synthesized-Trace (13.1\%$\rightarrow$8.8\%) and even on the Victim-Trace oracle (10.4\%), while Qwen stays roughly flat (28.9\%$\rightarrow$30.9\%). This likely reflects a mismatch between the inverted traces' math-flavored reasoning style and the coding-specific reasoning required by this benchmark, as well as some formatting differences between the reasoning supervision and the output format LCB expects.

\subsection{Stealing Reasoning Capabilities of a Black-Box, Commercial Model (GPT-5.4 mini)}
\label{sec:eval-gpt-stealing}

Table~\ref{tab:gpt_reasoning} evaluates capability stealing from a fully black-box commercial victim, GPT-5.4 mini, which exposes only $(x,b^\star,y)$. We train the inversion model offline with either a strong (R1) or weaker (R1-Weak) surrogate, then synthesize $\hat{t}$ for student fine-tuning.

\noindent\textbf{Inverted traces outperform answers and summaries.} Synthesized-Trace beats Answer-only and Summary+Answer on both students by large margins. With the R1-trained inverter, Qwen improves MATH500 from 68.4\% (Answer-only) to 76.0\% and JEEBench from 27.6\% to 43.7\%; Llama improves on MATH500 from 13.0\% to 52.4\% and JEEBench from 6.3\% to 19.9\%. Notably, Summary+Answer often underperforms Answer-only (e.g., Qwen JEEBench: 27.6\%$\rightarrow$1.6\%), suggesting that without inversion, summaries are not helpful for distillation.

\noindent\textbf{Inversion helps most when the surrogate is weak.} With the R1-Weak surrogate, Synthesized-Trace dominates Surrogate-Trace (Qwen JEEBench 19.7\%$\rightarrow$31.6\%, Llama 11.3\%$\rightarrow$16.8\%). With the strong R1 surrogate, Synthesized-Trace is competitive but can fall slightly below Surrogate-Trace (Qwen JEEBench 43.7\%$\rightarrow$43.7\%, Llama 33.2\%$\rightarrow$28.9\%).   A strong surrogate already supplies high-quality traces, while inversion must synthesize them indirectly from answers and summaries.

\begin{table*}[t]
  \centering
  \setlength{\belowcaptionskip}{6pt}
  \caption{\textbf{Downstream accuracy (\%) after fine-tuning when attacking GPT-5.4 mini.} }
  \label{tab:gpt_reasoning}
  \small
  \setlength{\tabcolsep}{4.4pt}
  \begin{tabular}{lll
  S[table-format=2.1] S[table-format=2.1]
  S[table-format=2.1] S[table-format=2.1]
  S[table-format=2.1] S[table-format=2.1]}
    \toprule
    \multirow{2.5}{*}{\textbf{Setting}} & \multirow{2.5}{*}{\textbf{Training Sources}} & \multirow{2.5}{*}{\textbf{Surrogate}}
    & \multicolumn{2}{c}{\textbf{MATH500}}
    & \multicolumn{2}{c}{\textbf{JEEBench}}
    & \multicolumn{2}{c}{\textbf{LiveCodeBench}} \\
    \cmidrule(lr){4-5} \cmidrule(lr){6-7} \cmidrule(lr){8-9}
    & & & \multicolumn{1}{c}{\textbf{Qwen}} & \multicolumn{1}{c}{\textbf{Llama}}
      & \multicolumn{1}{c}{\textbf{Qwen}} & \multicolumn{1}{c}{\textbf{Llama}}
      & \multicolumn{1}{c}{\textbf{Qwen}} & \multicolumn{1}{c}{\textbf{Llama}} \\
    \midrule
    -- & Answer-only & -- & 68.4 & 13.0 & 27.6 & 6.3 & 0.8 & 1.4 \\
    -- & Surrogate-Trace & R1 & 72.2 & 59.6 & 43.7 & 23.8 & 33.2 & 9.8 \\
    -- & Surrogate-Trace & R1-Weak & 63.2 & 48.8 & 19.7 & 11.3 & 22.9 & 9.1 \\
    \midrule
    Summary & Summary+Answer & -- & 19.8 & 16.4 & 1.6 & 5.5 & 0.0 & 0.4 \\
    Summary & \textbf{Synthesized-Trace (ours)} & \textbf{R1} & 76.0 & 52.4 & 43.7 & 19.9 & 28.9 & 10.0 \\
    Summary & \textbf{Synthesized-Trace (ours)} & \textbf{R1-Weak} & 73.2 & 48.3 & 31.6 & 16.8 & 25.6 & 10.2 \\
    \midrule
    No-sum. & \textbf{Synthesized-Trace (ours)} & \textbf{R1} & 71.8 & 47.2 & 41.6 & 16.6 & 27.7 & 9.7 \\
    \bottomrule
  \end{tabular}
  \vskip -0.1in
\end{table*}

As Table~\ref{tab:benchmark_results} (in Appendix~\ref{app:bench}) shows, R1 is comparable to GPT-5.4 mini (and even stronger on some benchmarks) . In this regime, Surrogate-Trace can be as effective as, or better than trace inversion.  We expect that inversion is most advantageous in the (common) setting when the victim is much stronger than the attacker's best surrogate, or when the victim outperforms the surrogate on the target task.

\begin{wrapfigure}{rt}{0.486\linewidth}
  \vspace{-0.50\baselineskip}
  \centering
  \includegraphics[width=1.0\linewidth]{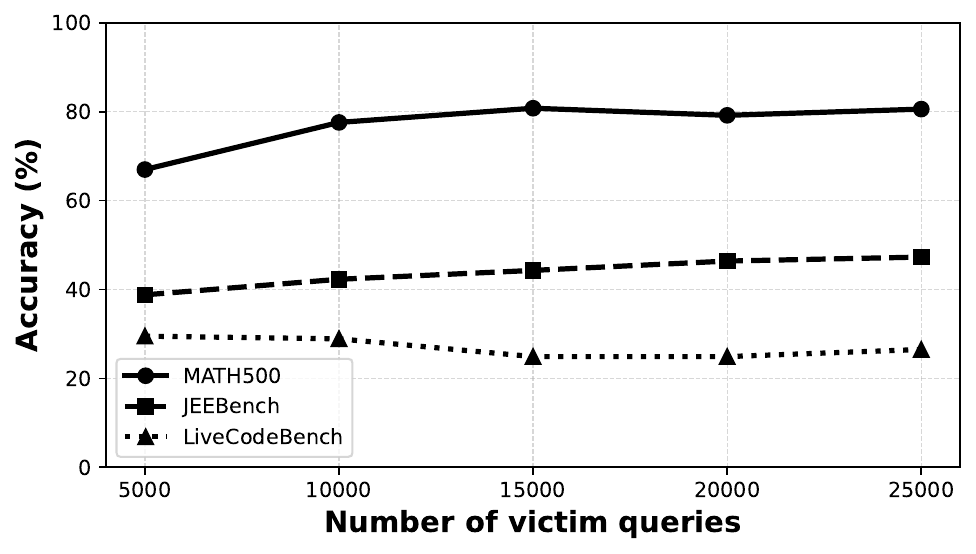}
  \caption{\textbf{Student accuracy vs. victim query budget.} Qwen trained on traces synthesized from GPT-5.4 mini using an R1-based inversion model.}
  \label{fig:query_scaling}
\end{wrapfigure}

\textbf{Inversion scales with the query budget.} Downstream accuracy improves consistently with the number of victim queries (Figure~\ref{fig:query_scaling}): scaling from 5k to 10k queries raises Qwen from 67.0\% to 77.6\% on MATH500 and from 38.8\% to 43.7\% on JEEBench; 15k queries reach 80.8\% and 44.5\%, respectively.  More queries expose the inversion model to diverse reasoning patterns, yielding higher-quality synthesized traces. LCB declines slightly with scale, consistent with the style/format mismatch discussed above. Our budgets ($\leq$25k) are far below production-scale distillation corpora (OpenThoughts: 114k; OpenThoughts3: 1.2M~\citep{guha2025openthoughtsdatarecipesreasoning}), scaling up is a direction for future work.

\textbf{Surrogate augmentation and domain-aware supervision.} Two practical adjustments can further improve the student, using the data already available to the attacker (Table~\ref{tab:gpt_further_improvements}). First, since R1 is itself a capable reasoning model comparable to GPT-5.4 mini, the R1 surrogate traces $\mathcal{D}_1$ generated in Stage~1 can be added to the distillation set ``for free.''  Training on $\{(x,\hat{t},y)\} \cup \mathcal{D}_1$ improves MATH500 (50.2\%$\rightarrow$51.6\%) and JEEBench (15.5\%$\rightarrow$19.9\%). Second, the attacker can \emph{domain-gate} the assistant target by including only the bare answer on code prompts. This sacrifices some accuracy on competition-style coding (LCB) but removes the redundant reasoning prefix that hurts the student elsewhere and sharply improves HumanEval+~\citep{liu2023your} (HE+ 20.7\%$\rightarrow$51.2\%), whose short self-contained Python functions are easier to solve with direct answers than LCB's long multi-constraint problems. The two adjustments compose: combining augmentation and gating yields the best overall trade-off (MATH500 57.6\%, JEEBench 21.6\%, HE+ 48.2\%). The broader point is that \textbf{inverted traces give the attacker a flexible training signal}: downstream behavior can be further tuned by mixing surrogate data or adjusting supervision format per domain without any additional queries to the victim.

\textbf{Capability stealing is economically feasible at moderate query budgets.}
Using current API pricing for \texttt{GPT-5.4 mini} ($\$0.75$/M input tokens; $\$4.50$/M output tokens), collecting 10k $(x,b^\star,y)$ queries costs \$173.28.  Inversion and student fine-tuning involve no GPT-5.4 mini queries.

\section{Defenses}
\label{sec:defenses}
Defenses against capability stealing aim to reduce the usefulness of a model’s outputs as fine-tuning data.  They include access control; output restrictions, e.g., query limits and truncation \citep{tramer2016stealing}; and perturbing outputs to reduce distillation fidelity while preserving task performance, including undistillable teachers, modified logits, and adaptive noise injection \citep{ma2021undistillable,ma2022stingy,chen2025queen}. 
Most of these methods apply to the internals (e.g., logits or intermediate representations) of image classifiers.  

Recent defenses of this type perturb reasoning to make it less transferable.  Antidistillation sampling \citep{savani2025antidistillation} and DOGe \citep{li2025doge} produce reasoning traces that yield correct answers but are intentionally hard to imitate, foiling chain-of-thought supervision.  These defenses target attacks that exploit the victim's reasoning.  In contrast, we operate in the realistic setting where the victim's reasoning is already hidden.  In practice, deployed models \citep{openai_reasoning_docs,gemini_thoughts_docs,anthropic_extended_thinking} expose only the final answers, sometimes with short reasoning summaries intended for user-facing explanation.  It is not known whether these summaries are specifically designed to foil distillation, but our experiments with fine-tuning student models on Answer-only or Summary+Answer show that direct distillation on them is not very effective. 

Even in this setting, trace inversion makes models vulnerable to capability stealing.  Without any access to the internal reasoning, operating only on the answers and summaries (if available), trace inversion synthesizes reasoning traces that enable student models to substantially improve their reasoning performance.  Trace inversion is completely agnostic to \emph{how} the answers and summaries were obtained, as long as they are correct, and thus immune to defenses that perturb internal reasoning. 

Complementary work explores watermarking generated text to enable post hoc detection of model use \citep{kirchenbauer2023watermark,gu2023learnability,sander2024watermarking}. While watermarking does not prevent capability stealing, it may support attribution; whether watermarks survive inversion and subsequent fine-tuning is an open question \citep{sander2024watermarking}.

\begin{table}[t]
  \centering
  \caption{\textbf{Further improvements on Synthesized-Trace (student=Llama, victim = GPT-5.4 mini).}}
  \label{tab:gpt_further_improvements}
  \small
  \begin{tabular}{l
  S[table-format=2.1] S[table-format=2.1]
  S[table-format=2.1] S[table-format=2.1]}
    \toprule
    \textbf{Training Sources} & {\textbf{MATH500}} & {\textbf{JEEBench}}  & {\textbf{LCB}} & {\textbf{HE+}}\\
    \midrule
    Synthesized-Trace & 50.2 & 15.5 & \textbf{10.0} & 20.7 \\
    $+$ surrogate augmentation  & 51.6 & 19.9 & 8.6 & 20.7 \\
    $+$ domain gating & 48.0 & 15.5 & 5.3& \textbf{51.2}  \\
    $+$ both & \textbf{57.6} & \textbf{21.6}& 6.1 & 48.2  \\
    \bottomrule
  \end{tabular}
\end{table}

\section{Discussion and Future Research}
\label{sec:discussion}
We demonstrated that hiding chains of thought may not protect reasoning capability from theft.  Even if a model exposes only a short reasoning summary (or no summary at all), an attacker-trained inversion model can use black-box access to synthesize long-form traces that are more effective than either the model's answers, or surrogate reasoning traces in teaching student models how to reason.

\emph{Fine-tuning effectiveness matters more than accurate reconstruction}.  For closed-source victims, we cannot tell whether our inverted traces accurately match their ``true'' reasoning traces.  It does not matter in practice, however,  as long as users can exploit their access to these closed models to improve the reasoning capabilities of their own models.

A key implication for defenses is that \emph{making exposed reasoning hard to imitate is not sufficient}, because an attacker can always ignore it and apply trace inversion to outputs alone. Obfuscated reasoning thus reduces transparency for users yet does not prevent capability stealing. 

Even stronger trace inversion is possible. We drew queries from OpenThoughts at a modest budget (at most 25K), compared to corpora such as OpenThoughts itself at 114K and OpenThoughts3 at 1.2M \citep{guha2025openthoughtsdatarecipesreasoning}. 
One natural direction is to scale up: larger query budgets, larger inversion models, and a broader mix of reasoning tasks. A complementary direction is to make inversion models more robust by training them on a variety of summary lengths and styles. Finally, trace synthesis can benefit from  verification (e.g., self-consistency for reasoning, execution for code, or tool-based checking for math) to filter or refine traces before student fine-tuning.

\section*{Acknowledgments.}
Supported in part by an Amazon Research Award, Google Academic Research Award, Google Cyber NYC Institutional Research Program, a research gift from Infosys, and NSF awards 2311521 and 2428949.

\section*{Impact Statement}
\label{sec:impact-statement}
This paper investigates whether hiding chains of thought (exposing only final answers or short reasoning summaries) prevents theft of reasoning capabilities. The sole purpose of this work is to evaluate the limits of existing protections and to motivate development of more robust defenses.

\bibliographystyle{abbrv}
\bibliography{main}

\newpage
\appendix
\section{Benchmark Accuracy
of Victim and Surrogate Models}
\label{app:bench}

We report zero-shot accuracy of R1, \texttt{R1-Distill} (R1-Weak), and GPT-5.4-mini on MATH500, JEEBench, and LiveCodeBench (\textbf{LCB}) in Table~\ref{tab:benchmark_results}.

\begin{table}[h]
  \centering
  \caption{\textbf{Benchmark accuracy (\%) of victims and surrogates.} }
  \label{tab:benchmark_results}
  \small
  \setlength{\tabcolsep}{8.5pt}
  \begin{tabular}{lccc}
    \toprule
    \textbf{Model} & \textbf{MATH500} & \textbf{JEEBench} & \textbf{LCB} \\
    \midrule
    R1          & 91.6 & 87.1 & 74.9 \\
    R1-Distill  & 81.4 & 32.6 & 27.6 \\
    GPT-5.4-mini  & 91.4 & 85.1 & 66.9 \\
    \bottomrule
  \end{tabular}
  \vskip -0.1in
\end{table}

\section{Prompts for Compression and Zero-shot Trace Inversion}

\phantomsection
\label{box:summarization_prompt}
\begin{tcolorbox}[breakable, pad at break*=1mm, title=Reasoning Summary Compression Prompt]

\textbf{System prompt.}\\
You are summarizing a long chain-of-thought trace into a short, first-person ``inner-monologue recap'' that mirrors the style of GPT-5 mini's internal reasoning summaries.\\

Given a \texttt{<think>...</think>} trace, produce a recap with these properties:\\

\textbf{Structure.} Write 3 to 6 short sections. Each section begins with a short bold markdown header on its own line, like \texttt{**Setting up the integral**} or \texttt{**Checking the edge case**}, and is followed by one short paragraph (2--5 sentences). Do NOT use numbered lists (\texttt{1.}, \texttt{2.}) and do NOT use bullet points (\texttt{-}, \texttt{*}). The whole recap is just headers + prose.\\

\textbf{Voice.} First person, present tense, as if the model is thinking aloud: ``I need to\ldots'', ``I'll check\ldots'', ``I'm now realizing\ldots'', ``Let me verify\ldots''. Keep the tone tentative and exploratory, not textbook.\\

\textbf{Content.} Each section should capture one meaningful move in the reasoning --- a clarification of what is being asked, a key derivation or substitution, a pivot after a failed attempt, a sanity check, or the final consolidation. Skip filler, restatement, and purely mechanical arithmetic. Preserve the logical order of the trace.\\

\textbf{Length.} Aim for roughly 600--900 tokens total --- long enough that each section develops a real idea, not just a one-line gesture. If the input trace is very short, produce fewer sections rather than padding; if it is very long, prefer adding depth to each section over adding more sections.\\

\textbf{Math formatting.} Use inline LaTeX (\texttt{\textbackslash frac}, \texttt{\textbackslash sqrt}, \texttt{\textbackslash boxed}, etc.) where the original trace used math. Do not restate the final boxed answer unless the reasoning naturally concludes with it.\\

Do not add meta-commentary about following instructions, apologize, or mention that you are summarizing. Just produce the recap.\\

\vspace{-10pt}
\rule{\linewidth}{0.3pt}\\
\vspace{-10pt}

The prompt additionally includes two few-shot exemplars of the target style (one algebra / number theory and one geometry, $\approx$600 tokens each), drawn from GPT-5-mini's own summaries. They are omitted from this box for space; the full prompt including exemplars is released with our code.\\

\vspace{-10pt}
\rule{\linewidth}{0.3pt}\\
\vspace{-10pt}

\textbf{User prompt.}\\
Summarize this thinking process as a first-person inner-monologue recap:\\
\texttt{<think>\{thinking\_content\}</think>}

\end{tcolorbox}

\phantomsection
\label{box:zero_shot_inversion_summary}
\begin{tcolorbox}[breakable, pad at break*=1mm, title=Zero-shot Inversion Prompt (With Summaries)]
You are a language model that reconstructs full internal reasoning traces from high-level bubble summaries.\\

You will be given:\\
- A problem **input** (e.g., a math or logic problem)\\
- A final **output** or solution\\
- A list of numbered **reasoning bubbles**, where each bubble summarizes one key insight, step, or decision made during the problem-solving process\\

These bubbles are **condensed summaries** of what was originally a much longer, richer internal thought process.\\
Your task is to reconstruct that full process.\\

Below are high-level bubble summaries representing condensed thoughts or decisions. Your task is to reconstruct the full thinking trace that might have led to each summary. For each bubble, expand it into a **detailed internal monologue or reasoning chain**, showing how one idea leads to the next.\\

Include:\\
- Assumptions and background intuitions\\
- Intermediate steps, definitions, and subcases\\
- Natural questions or doubts raised during reasoning\\
- Alternatives that were considered and rejected\\
- Transitions that make the reasoning coherent and plausible\\

Use informal, introspective language — as if the person is thinking out loud. Add math expressions in \LaTeX where appropriate.\\

Do **not** invent new reasoning steps outside the bubbles. Use the **input** and **output** only for context and consistency. Your goal is to **flesh out the bubbles**, not to re-solve the problem from scratch.\\

The full trace should:\\
- Be logically consistent and cohesive from start to finish\\
- Sound like a realistic thought process that could plausibly result in the given answer\\
- Span multiple paragraphs per bubble and up to 20,000 characters overall if needed\\
- Be output as one continuous trace, wrapped in `<think>...</think>` tags\\

You are not summarizing the bubbles. You are recovering the internal narrative that *generated* them.\\

The original problem input is: \texttt{\{user\_prompt\}}\\
The final answer is: \texttt{\{assistant\_answer\}}\\
Transform this thinking bubbles into clear full reasoning traces: \texttt{\{reasoning\_summary\}}\\
\textbf{Generate full reasoning traces:}

\end{tcolorbox}

\newpage
\phantomsection
\label{box:zero_shot_inversion_no_summary}
\begin{tcolorbox}[breakable, pad at break*=1mm, title=Zero-shot Inversion Prompt (No Summaries)] 
You are a language model that reconstructs full internal reasoning traces from only an **input** (e.g., a math or logic problem) and a corresponding **output** (final solution or answer). \\

You will be given:\\
- A problem **input**\\
- A final **output** or solution\\

Your task is to reconstruct the full internal reasoning process that could plausibly connect the input to the output. This should be a long, detailed, introspective trace, not a short summary.\\

Guidelines for the reasoning trace:\\
- Write in the style of an informal, introspective monologue, as if the person is thinking out loud.\\
- Include assumptions, intuitions, and background facts as they arise naturally.\\
- Show intermediate steps, calculations, logical deductions, definitions, and subcases.\\
- Raise natural questions or doubts during reasoning, and explain how they are resolved.\\
- Explore alternative approaches, even ones that are discarded, and explain why.\\
- Make transitions clear so the reasoning feels like a coherent train of thought.\\
- Use \LaTeX for math expressions where helpful.\\
- Do not introduce new information inconsistent with the input or output.\\
- The goal is depth, not brevity: expand ideas fully, elaborate with multiple paragraphs, and let the reasoning unfold gradually.\\
- The output should only appear at the end, after the reasoning is complete.\\

The full trace should:\\
- Wrapped in <think>...</think> tags.\\
- Be logically consistent and cohesive from start to finish.\\
- Sound like a realistic thought process that could plausibly result in the given answer.\\
- Span thousands of tokens if needed.\\

The original problem input is: \texttt{\{user\_prompt\}}\\
The final answer is: \texttt{\{assistant\_answer\}}\\
\textbf{Generate full reasoning traces:}
\end{tcolorbox}

\newpage
\section{Examples of Synthesized Reasoning Traces}

\begin{figure*}[!ht]  
  \centering    \includegraphics[width=\linewidth]{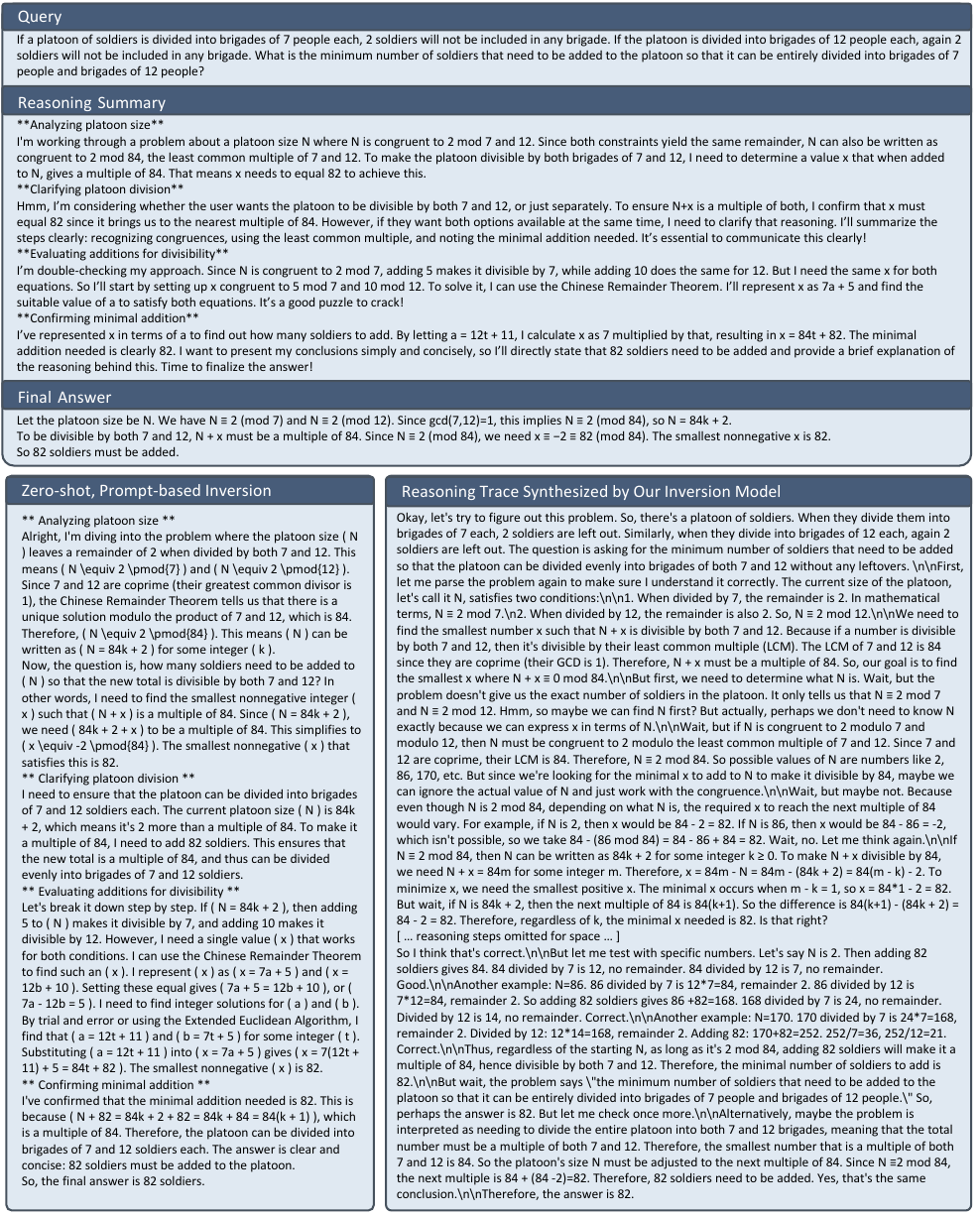}
  \caption{\textbf{Trace inversion against R1.} Zero-shot inversion versus trace inversion with \textbf{R1-Distill} as the surrogate model.}
  \vskip -0.2in
  \label{fig:example_2}
\end{figure*}

\begin{figure*}[!ht]
  \centering    \includegraphics[width=\linewidth]{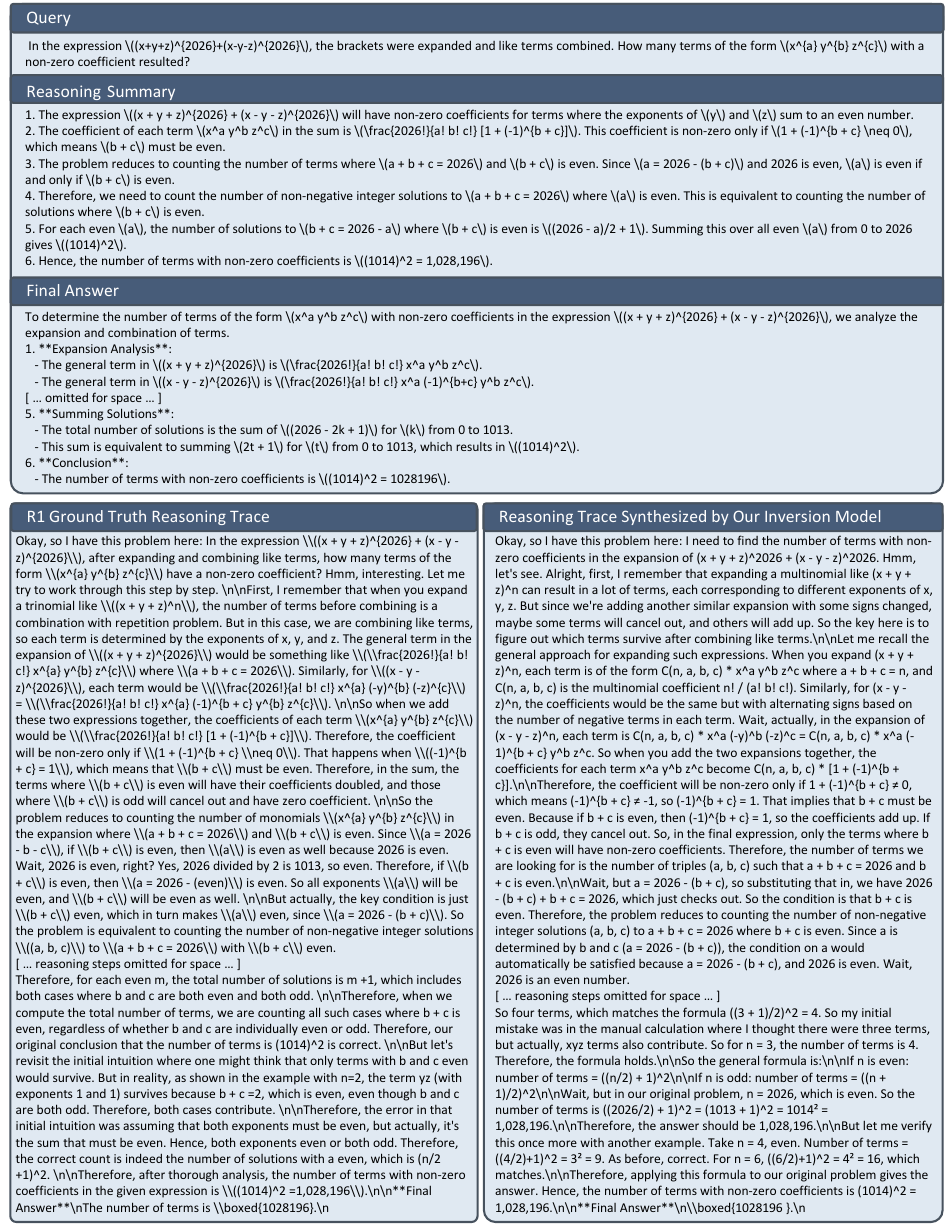}
  \caption{\textbf{Trace inversion against R1.} Ground-truth reasoning trace versus a trace synthesized with \texttt{R1-Distill} as the surrogate model.}
  \label{fig:example_3}
\end{figure*}

\clearpage
\section{Synthesized Traces Can Be Better Than Ground-Truth Traces}
\definecolor{qbg}{HTML}{E0E9F1}
\definecolor{qframe}{HTML}{475B78}
\definecolor{invbg}{HTML}{E0E9F1}
\definecolor{invframe}{HTML}{475B78}
\definecolor{r1bg}{HTML}{E0E9F1}
\definecolor{r1frame}{HTML}{475B78}
\definecolor{boxbrand}{HTML}{475B78}
\definecolor{hlbrand}{HTML}{C62828}

\begin{figure*}[h]
  \centering

\begin{tcolorbox}[
    colback=qbg, colframe=qframe,
    title={\small\textbf{Query}},
    fonttitle=\sffamily, fontupper=\small,
    boxrule=0.6pt, arc=2pt, left=4pt, right=4pt, top=2pt, bottom=2pt,
  ]
  The positive reals $x$ and $y$ satisfy $x^3 + y^3 = x - y$. Show that $x^2 + y^2 < 1$.
\end{tcolorbox}

\begin{minipage}[t]{0.485\textwidth}
  \begin{tcolorbox}[
    colback=r1bg, colframe=r1frame,
    title={\small\textbf{R1 Ground-Truth Reasoning Trace}},
    fonttitle=\sffamily, fontupper=\footnotesize,
    boxrule=0.6pt, arc=2pt, left=4pt, right=4pt, top=2pt, bottom=2pt,
  ]
  \ldots{} \textcolor{hlbrand}{\textit{Wait, but we already have $x - y = (x + y)(S - P)$ from the equation above.}} Wait, let's step back.\ldots

  Since $f$ is convex for $a > 0$, by Jensen's inequality: $(x^3 + y^3)/2 \geq ((x + y)/2)^3$. \textcolor{hlbrand}{\textit{But not sure if helpful here.}}\ldots

  Since $x > y$, we can write $x = y + k$ where $k > 0$. Substitute into the equation: $(y+k)^3 + y^3 = k$. Expanding: $2y^3 + 3y^2 k + 3yk^2 + k^3 = k$. \textcolor{hlbrand}{\textit{But this seems complicated. Maybe not the best approach.}}\ldots

  The inequality becomes $xy < 2y/(x+y)$. Since $y > 0$, we can divide both sides by $y$: $x < 2/(x+y)$. Multiply both sides by $(x+y)$: $x(x + y) < 2$. So $x^2 + xy < 2$. \textcolor{hlbrand}{\textit{But we need to relate this to our previous expressions. Hmm, not sure if helpful.}}\ldots
  \end{tcolorbox}
\end{minipage}
\hfill
\begin{minipage}[t]{0.485\textwidth}
  \begin{tcolorbox}[
    colback=invbg, colframe=invframe,
    title={\small\textbf{Synthesized Reasoning Trace}},
    fonttitle=\sffamily, fontupper=\footnotesize,
    boxrule=0.6pt, arc=2pt, left=4pt, right=4pt, top=2pt, bottom=2pt,
  ]
  Another idea: \ldots\ $x^2 - xy + y^2 = (x-y)/(x+y)$. But $x^2 + y^2 = (x^2 - xy + y^2) + xy$, so $x^2 + y^2 = (x-y)/(x+y) + xy$.

  Let's denote $t = y/x$, since $x > y$, so $t$ is a positive real number less than 1. \ldots\ substitute into the equation $x^3 + y^3 = x - y$: $x^3(1 + t^3) = x(1 - t)$. Therefore, $x^2 = (1 - t)/(1 + t^3)$. Now, $x^2 + y^2 = x^2(1 + t^2) = (1 - t)(1 + t^2)/(1 + t^3)$.

  \ldots\ Expand the left-hand side: $1 + t^2 - t - t^3 < 1 + t^3$. \ldots\ Factor out a $t$: $t(2t^2 - t + 1) > 0$. Discriminant $D = (-1)^2 - 4 \times 2 \times 1 = 1 - 8 = -7$. Since the discriminant is negative, the quadratic $2t^2 - t + 1$ has no real roots and since the coefficient of $t^2$ is positive, it is always positive. \ldots\ we conclude that $x^2 + y^2 < 1$.''\ldots
  \end{tcolorbox}
\end{minipage}
  \caption{\textbf{R1 ground-truth trace vs.\ synthesized trace with R1 as the surrogate.} \textbf{Top:} the query. \textbf{Bottom left:} excerpt from R1's ground-truth trace, which tries and abandons several approaches.
  \textbf{Bottom right:} matching excerpt from the trace synthesized by our inversion model (trained with R1 as the surrogate), with more direct reasoning.}
  \label{fig:r1_vs_inverter_backtrack}
\end{figure*}


\end{document}